\documentclass[final,3p,times,twocolumn]{elsarticle}
 \biboptions{comma,sort&compress}
 \usepackage{graphicx}
  \usepackage{epsfig}

\usepackage{amsmath}
\usepackage {amsfonts,amssymb,amstext}

\def\nin{\noindent}
\def\beq{\begin{equation}}
\def\eeq{\end{equation}}
\def\bea{\begin{eqnarray}}
\def\eea{\end{eqnarray}}
\def\nnb{\nonumber}

\newcommand{\cF}{{\cal F}}

\newcommand{\cO}{{\cal O}}

\newcommand{\setl}{\setlength\arraycolsep{2pt}} 
\newcommand{\Nc}{\mbox{${\rm N_c}$}}

\newcommand{\MeV}{\mbox{\rm MeV}}

\journal{Nuc. Phys. (Proc. Suppl.)}

\begin{document}

\begin{frontmatter}



\title{\Large Update of the Electron and Muon \boldmath${\rm g}$-Factors}

 \author{Eduardo de Rafael\corref{label2}}
 \address{Aix-Marseille Universit\'e, CNRS, CPT, UMR 7332, 13288 Marseille, France\\
 Universit\'e de Toulon, CNRS, CPT, UMR 7322, 83957 La Garde, France}     
\ead{EdeR@cpt.univ-mrs.fr}


\begin{abstract}
\noindent 
I give an overview of the different contributions to the electron and muon anomalous magnetic moments in the Standard Model. Special emphasis is given  to recent QED results as well as to the hadronic light-by-light scattering contribution to the muon anomaly.

\end{abstract}


\end{frontmatter}


\section{Introduction}
\nin
The $g$--factors of the leptons $e$, $\mu$ and $\tau$ are the dimensionless parameters which relate their spin $\vec{s}$ to their magnetic moments $\vec{\mu}_l$:
\beq
\vec{\mu}_{l}=g_{l}\frac{e\hbar}{2m_{l}c}\vec{s}\,.
\eeq
The anomalous magnetic moment $a_l$ (anomaly for short), is the correction to the Dirac value $g_l =2$:
\beq
 \underbrace{g_{l}=2}_{\it Dirac}\left(1+\underbrace{a_{l}=\frac{\alpha}{2\pi} +\cdots}_{anomaly}\right)\,,\quad l=e\,, \mu\,, \tau\,.
\eeq
The lowest order contribution to the anomaly in powers of the fine structure constant $\alpha$, the Schwinger term $\frac{\alpha}{2\pi}$, is the same for the three leptons. Higher order contributions become sensitive to the masses of the virtual particles exchanged and obey the following pattern:
\begin{itemize}
	\item Heavy virtual particles with $M>m_l$ decouple as powers of $\frac{m_l^2}{M^2}$ modulated by powers of $\log\left(\frac{M^2}{m_l^2} \right)$ factors.
	\item Light virtual particles: $e$ for $l=\mu\,,\tau$ and $e\,,\mu$ for $l=\tau$, bring in powers  of log enhancements of the mass ratios.
	\item Virtual Hadronic Interactions and Electroweak Interactions become relevant at the present degree of experimental accuracy.
	\item Because of the relatively large value of the muon mass, the muon anomaly with its high precision measurement provides a sensitive probe for new fundamental physics.
	\item The tau anomaly is too poorly known experimentally to provide a test of the Standard Model. 
\end{itemize}
\section{Electron Anomaly and Fine Structure Constant}
\nin
The best determination of the electron anomaly comes from experiments by the Harvard group~{\cite{Gabetal06}} with the result:
\beq\label{eq:harvard}
a_{e}(\rm exp.) =  1~159~652~180.73~(0.28)\times 10^{-12}\,,
\eeq
which is a $0.24~{\rm ppb}$ precision measurement! The Quantum Electrodynamics (QED) contributions of the massless class of Feynman diagrams (diagrams with virtual photons only as well as  with virtual photons and fermion loops of the same flavour as the external particle, the electron in our case) have been evaluated up to tenth order
\beq
a_{e}({\rm QED-massless})=\sum_{n=1}^{n=5} A_{1}^{(2n)}\left(\frac{\alpha}{\pi} \right)^{n}
\eeq
with the following results:

{\setl
\bea
A_{1}^{(2)} & = & +~0.5\\
A_{1}^{(4)} & = & -~0.328~478~965~579~193~\cdots \\
A_{1}^{(6)} & = & +~1.181~241~456~\cdots \\
A_{1}^{(8)}& = & -~1.9106~(20) \\
A_{1}^{(10)} & = & +~9.16~(58)\,.
\eea}

\nin
The coefficient $A_{1}^{(2)}$ was first calculated by Schwinger in 1948~{\cite{Sch49}}. Ten years later Petermann and Sommerfield, independently, calculated $A_{1}^{(4)}$~{\cite{Pe57,So57}}. The sixth order term $A_{1}^{(6)}$ is also known {\it analytically} thanks to the de\-di\-cated work of Laporta and Remiddi~\cite{LaRe96} . The dots in $A_{1}^{(4)}$ and $A_{1}^{(6)}$ indicate that these are known numbers to any desirable accuracy. The eigth and tenth order coefficients $A_{1}^{(8)}$ and $A_{1}^{(10)}$ have been calculated {\it numerically} by Kinoshita and collaborators~\cite{Kino07,AHKN12}. The calculation of $A_{1}^{(10)}$ in particular involves 12672 Feynman diagrams! This is an extraordinary achievement after many years of dedicated effort. Notice the alternative signs for the coefficients $A_{1}^{(2n)}$: positive for $n$-odd and negative for $n$-even; an interesting pattern for which we have no simple explanation.

With $\frac{m_{e}^{2}}{m_{\mu}^2}$ and $\frac{m_{e}^{2}}{m_{\tau}^2}$ power corrections incorporated, as well as Hadronic Vacuum Polarization ($\sim 2\times 10^{-12}$), Hadronic Light-by-Light Scattering~\footnote{See the estimate discussed in ref.~{\cite{PdeRV10}}.}($\sim 3\times 10^{-14}$) and Electroweak ($\sim 3\times 10^{-14}$) corrections incorporated we have at present the following theoretical prediction:
{\setl
\bea
a_{e}({\rm SM}) & = & 1~159~652~181.82\nnb\\
 & & \!\!\!\!\!\!\!\!\underbrace{(6)}_{\rm\footnotesize eighth}\underbrace{(4)}_{\rm\footnotesize tenth}\underbrace{(2)}_{\rm\footnotesize H-EW}\underbrace{(78)}_{~\alpha^{-1}({\rm Rb})}\times 10^{-12}\,,
\eea}

\nin
which has a $0.67~{\rm ppb}$ precision. Notice that the largest error here comes from inserting the Atomic Physics determination of $\alpha$~\cite{Ry11}. One can proceed otherwise: extract the value of $\alpha$ from the comparison between the predicted value of $a_{e}({\rm SM})$ and the Harvard measurement in Eq.~{\ref{eq:harvard}}, with the result:

{\setl
\bea
\alpha^{-1}(a_e) & = & 137.035~999~1736 \nnb \\
& & \!\!\!\!\!\!\!\! \underbrace{(68)}_{\rm\footnotesize eighth}\underbrace{(46)}_{\rm\footnotesize tenth}\underbrace{(26)}_{\rm\footnotesize H-EW}\underbrace{(331)}_{\rm Harvard})\,,
\eea}

\nin
which represents a  $0.25~{\rm ppb}$ precision measurement and becomes at present the reference value of $\alpha$ for all the other QED observables ($a_{\mu}$ in particular).
\section{Muon Anomaly}
\nin
The present experimental world average determination of the muon anomaly is dominated by the latest BNL experiment of the E821 collaboration~\cite{g2col} with the result:
\beq\label{eq:bnl}
a_{\mu}(\rm\small E821-BNL)= 116~592~089 (54)_{\rm\tiny stat} (33)_{\rm\tiny syst}\times 10^{-11}\,,
\eeq
which is a $0.54~{\rm ppm}$  precision measurement.

At the level of the experimental accuracy, the QED contributions to $a_{\mu}$ from photons and leptons alone are very well known. The results, due to the work of many people~\footnote{For a recent review article  where earlier references can be found see~\cite{MdeRRS12}.\label{foot:review}}, are summarized in  Table~1 below:
\begin{center}
{\bf Table~1: QED Contributions (Leptons)}\\
\{$\alpha^{-1}=137.035~999~1736~(342)~[0.25~{\rm ppb}]$\}
{\small
\begin{tabular}{|c|r|} \hline \hline {\sc\small Contribution} &
{\sc\small  Result in Powers of $\frac{\alpha}{\pi}$} 
\\ \hline \hline   
$a_{\mu}^{(2)}$ & $0.5\left(\frac{\alpha}{\pi}\right)\ $ \\  
\hline 
$a_{\mu}^{(4)}(\rm total)$ & $0.765~857~425 ~(17)\left(\frac{\alpha}{\pi}\right)^2 $ \\
\hline 
$a_{\mu}^{(6)}(\rm total)$ & $24.050~509~96~(32)\left(\frac{\alpha}{\pi}\right)^3$ \\
\hline 
$a_{\mu}^{(8)}(\rm total)$  & $130.879~6~(63)\left(\frac{\alpha}{\pi}\right)^4$  \\ \hline 
$a_{\mu}^{(10)}(\rm total)$ & $753.29~(1.04)\left(\frac{\alpha}{\pi}\right)^5$  
 \\ 
\hline 
 $a_{\mu}(\rm QED)$ &  
  $116~584~718.845~(37)\times 10^{-11}$  \\
 \hline\hline
\end{tabular}}
\end{center}
The tenth order result  $a_{\mu}^{(10)}(\rm total)$  in  Table~1 is the one recently published by Kinoshita {\it et~al}~\cite{Kinalmu12}. Again this is the result of very long impressive calculations improved during a long period of time.

The Standard Model contributions to $a_{\mu}$ with the Hadronic Vacuum Polarization of lowest order (HVP(lo)) and higher order (HVP(ho), as well as from  Hadronic Light-by-Light Scattering (HLbyL) and Electroweak (EW) effects incorporated are summarized in Table~2 below:

\begin{center}
{\bf Table~2: Standard Model Contributions$^{\ref{foot:review}}$}
\begin{tabular}{lr}
\hline \hline {\sc\small Contribution} &
{\sc\small  Result in $10^{-11}$ units } 
\\ \hline     
QED (leptons) & $116~584~718.85\pm  0.04$ \\
HVP(lo)[$e^+ e^-$] & 
$6~923\pm { 42}$ \\
HVP(ho) &  $-98.4\pm 0.7$ \\
HLbyL & $105\pm { 26}$ \\
EW & $153\pm 1$ \\ 
 \hline
 Total SM & $116~591~801\pm {49}$\\
  \hline\hline\ 
\end{tabular}
\end{center}

The persistent  $3.6~\sigma$ discrepancy between the total SM result and the experimental value in Eq.~{\ref{eq:bnl}} deserves attention. In that respect,
one observes that the largest errors in Table~2  come from the HVP and the HLbyL contributions. The size of these errors is of special concern in view of the future experiments at FNAL in the USA and at JPARC in Japan which plan to reduce the present experimental error from $0.54~{\rm ppm}$ to $0.14~{\rm ppm}$. We examine these contributions in the next section.
\section{Hadronic Vacuum Polarization and Hadronic Light-by-Light Scattering Contributions}
\begin{figure}[h]
\begin{center}
\includegraphics[width=0.25\textwidth]{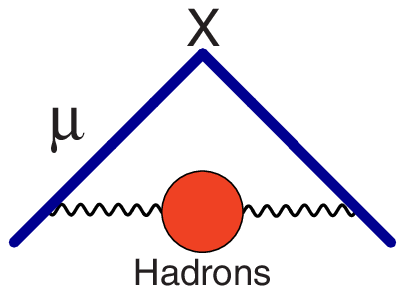}
\bf\caption{\label{fig:hvp}}
{\it  Hadronic Vacuum Polarization Contribution.}
\end{center}
\end{figure}

\nin
The lowest order HVP contribution to the muon anomaly, shown in Fig.~{\ref{fig:hvp}}, has a well known integral representation in terms of the $e^+ e^-$ one photon annihilation cross--section  into hadrons $\sigma_{\rm H}(t)$, where $t$ denotes the total CM energy squared: 
\beq
a_{\mu}^{({\rm HVP})}\!=\!\frac{1}{4\pi^2}\int_{4 m_{\pi}^2}^\infty\! dt\int_0^1 \! dx\frac{x^2 (1-x)}{x^2 +\frac{t}{m_{\mu}^2}(1-x)} \sigma_{\rm H}(t)\,.
\eeq

\nin
This integral is in fact dominated by the contributions from the low-$t$ region. Its determination has been im\-pro\-ving through the years thanks to the advent of more and more refined measurements; the latest coming from the BaBar and Belle facilities. The error here is likely to be reduced in the near future.

\begin{figure}[h]
\begin{center}
\includegraphics[width=0.40\textwidth]{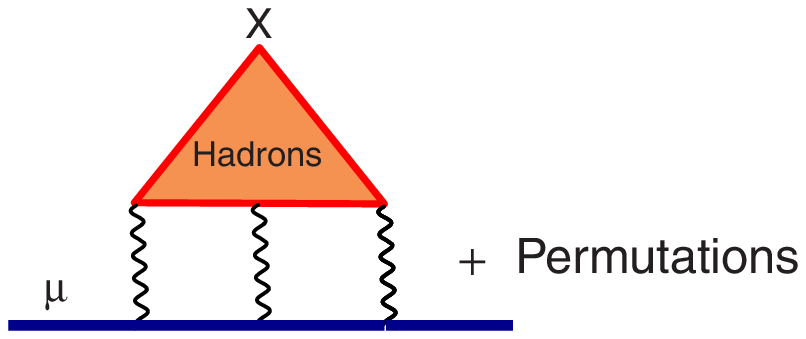}
\bf\caption{\label{fig:hlbyl}}
{\it  Hadronic Light-by-Light Scattering Contribution.}
\end{center}
\end{figure}

Contrary to the HVP contribution, the Hadronic Light--by--Light Scattering contribution (HLbyL) shown in Fig.~{\ref{fig:hlbyl}} cannot be written as an integral over ex\-pe\-ri\-mentally accesible observables. 
The calculation from theory requires the knowledge of Quantum Chromodynamics (QCD) contributions at all energy scales, something which we don't know how to do at present except via numerical lattice QCD evaluations which, unfortunately, are difficult to implement in this case~\footnote{There are, however, promising lattice projects under study, see e.g. ref.~{\cite{Blum12}}.}. 

Things, however, are not as hopeless as all that. We know some important constraints from QCD which the hadronic models one has to resort to, to evaluate this contribution, have to obey. 
\begin{itemize}
	\item {\it Chiral Limit and Large--$\Nc$ limit of QCD}
	
	It is well known that the hadronic realization of QCD in the sector of the  light $u\,,d\,,s$ quarks is governed by the phenomena of {\it spontaneous chiral symmetry breaking} and {\it confinement}. This implies the existence of a {\it Mass Gap} between the pseudoscalar ({\it Goldstone--like}) particles and the other hadronic particles (the $\rho$ being the lowest one in Nature). In the limit where this {\it Mass Gap} is considered to be large, and to leading order in the $1/\Nc$--expansion ($\Nc$ is the number of colours and the expansion in question is a topological expansion of phenomenological relevance), the contribution of  the HLbyL scattering to the muon anomaly is dominated by the diagrams shown in Fig.~{\ref{fig:hlbylpi}}
\begin{figure}[h]
\begin{center}
\includegraphics[width=0.45\textwidth]{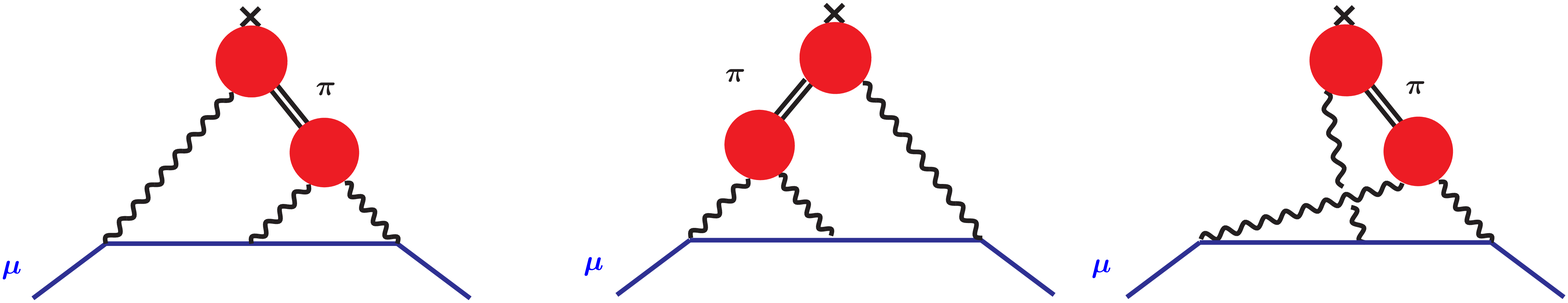}
\bf\caption{\label{fig:hlbylpi}}
{\it  Dominant HLbyL Scattering Contribution.}
\end{center}
\end{figure}
where only one pion propagator appears. The leading result in this limit is known analytically~\cite{KNPdeR02}:
	
{\setl
\bea\label{eq:knpder}
a^{(\rm HLbyL)}_{\mu}(\pi^{0}) & = & \Big(\frac{\alpha}{\pi}\Big)^{3}\ \Nc^2 \frac{m_{\mu}^{2}}{48\pi^2 f_{\pi}^{2}}
\ \Big[ \log^2\frac{M_{\rho}}{m_{\pi}}\nnb\\ 
&  & +\  \cO\Big(\log\frac{M_{\rho}}{m_{\pi}}\Big)+\cO(1)\Big]\,.
\eea}

\nin
Notice that the leading order approximation in the $1/\Nc$--expansion is here necessary because it selects the one--pion exchange contribution; pion loops can only appear at the next--to--leading level. 

This result is phenomenologically relevant because it can be used as a check of Hadronic Model Calculations: letting the hadronic masses of a model become very large one must find an answer compatible with the analytic result in Eq.~{\ref{eq:knpder}} when $M_{\rho}$ is also taken to be large. The fact that the HLbyL contribution in this limit is positive was crucial in fixing the overall sign of $a^{(\rm HLbyL)}_{\mu}$~\cite{KN02}. Numerically, for the physical values of the constants in Eq.~{\ref{eq:knpder}} one finds
\beq
\Big(\frac{\alpha}{\pi}\Big)^{3}\ \Nc^2 \frac{m_{\mu}^{2}}{48\pi^2 f_{\pi}^{2}}
\ \log^2\frac{M_{\rho}}{m_{\pi}}= 95\times 10^{-11}\,,
\eeq
which is within the ballpark of the hadronic model determinations. The $\rho$-mass, however, is too close to the $\pi$-mass to take seriously this number which may have large corrections from the $\cO\Big(\log\frac{m_{\rho}}{m_{\pi}}\Big)$ and $\cO(1)$ terms. In fact, it is in principle possible to fix the coefficient of the $\cO\Big(\log\frac{m_{\rho}}{m_{\pi}}\Big)$ term. It can be shown~\cite{Ny02,RMW02} that the unknown contribution to this coefficient is related to the $\pi^0 \rightarrow e^+ e^-$ decay rate, albeit with the radiative corrections included. The $\cO(1)$ term is, however,  not fixed by symmetry restrictions which  limits the use of Eq.~{\ref{eq:knpder}} at the  level of the required accuracy.

\item{\it Short--Distance Constraint from the OPE in QCD}

The constraint in question comes from a clever observation by Melnikov and Vainshtein~\cite{MV04}. The three momenta $k_1 \,,k_2 \,, k_3$ in the light--by--light subdiagram of the HLbyL scattering contribution to the muon anomaly (see Fig~\ref{fig:mv} below)
\begin{figure}[h]
\begin{center}
\includegraphics[width=0.4\textwidth]{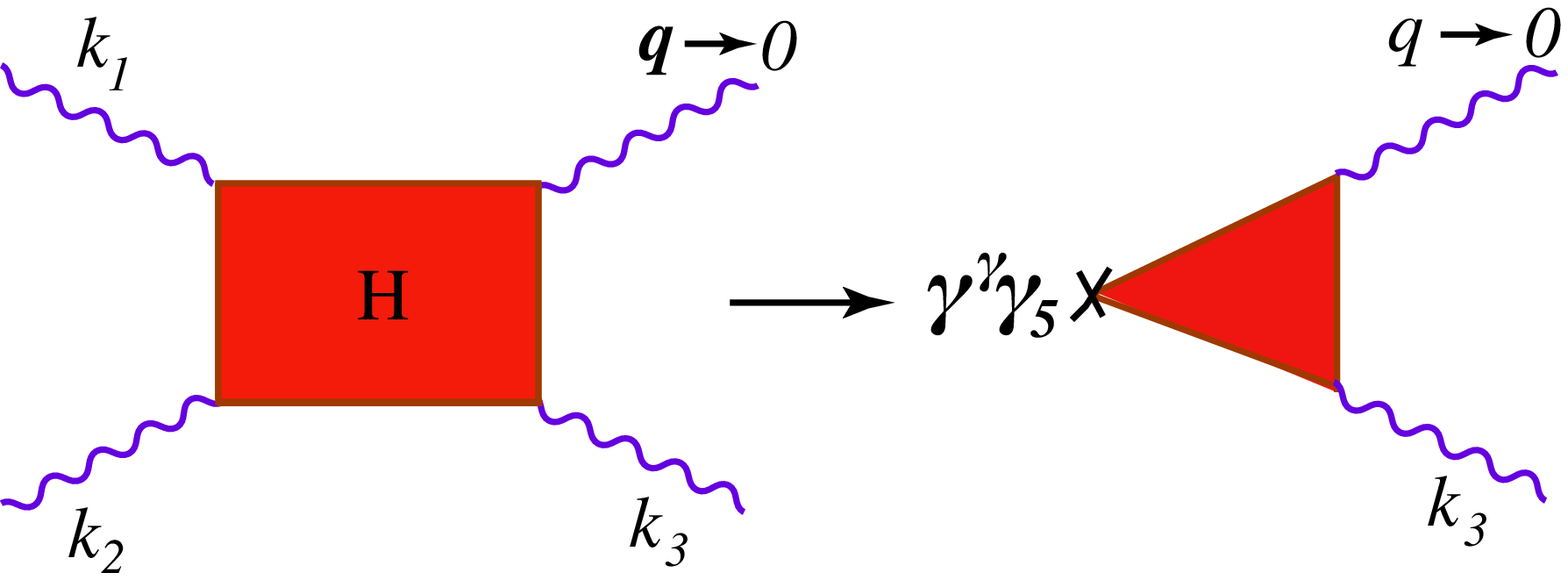}
\bf\caption{\label{fig:mv}}
{\it  The OPE Constraint of Melnikov and Vainshtein.}
\end{center}
\end{figure}
form a tri\-angle. When $k_{1}^{2}\approx k_{2}^{2} \gg k_{3}^{2}$\,, and  $k_{1}^{2}\approx k_{2}^{2}\gg m_{\rho}^{2}$ in this triangle one can apply the OPE in the two  vector currents which carry hard momenta with the result:

{\setl
\bea\lefteqn{
\int d^4x_1 \int d^4x_2  {\rm e}^{-ik_1\cdot x_1 -ik_2\cdot x_2} 
 J_{\nu}(x_1) J_{\rho}(x_2)   = \nnb} \\ & & 
 \frac{2\epsilon_{\nu\rho\delta\gamma}{\hat k}^{\delta}}{\hat k^{2}}\!\! \int d^4z  {\rm e}^{-ik_{3}\cdot z} J_{5}^{\gamma}(z)+{\cal O}\bigg(\frac{1}{{\hat k}^{3}}\bigg)\,,
\eea}

\nin
where $j_{5}^{\gamma}=\sum_{q} Q_{q}^{2}\,\bar q \gamma^{\gamma}\gamma_{5}q$ is the axial current 
with the different flavors weighted by squares of their electric charges and $\hat k=(k_{1}-k_{2})/2\approx k_{1}\approx -k_{2}$\,.  As illustrated in Fig.~\ref{fig:mv} this OPE reduces the HLbyL amplitude, in 
the special kinematics under consideration, to the AVV triangle amplitude which is an object for which we have a much better theoretical insight.
This observation has interesting phenomenological implications:
\begin{itemize}
\item At large $k_{1,2}$ the Pseudoscalar (and Axial-Vector) exchanges dominate.

\item The AVV limit also implies that the  $\cF_{\pi^0 \gamma^* \gamma^*} (k^2 , k^2 )$ form factor at the vertices of Fig~\ref{fig:hlbylpi} must fall as $1/k^2$.
\end{itemize}
\end{itemize}

Unfortunately, the two asymptotic QCD constraints discussed above are not sufficient for a full model independent evaluation of $a^{(\rm HLbyL)}_{\mu}$. This explains the relatively large error of $\pm 26\times 10^{-11}$ for this contribution in Table~2 above.

Most of the last decade calculations of $a^{(\rm HLbyL)}_{\mu}$ found in the literature~$^\text{\ref{foot:review}}$ are compatible with the QCD chiral constraints and the large--$\Nc$ limit discussed above. They all incorporate the $\pi^0$--exchange contribution modulated by $\pi^0 \gamma^* \gamma^*$  form factors correctly normalized to the Adler, Bell--Jackiw  point--like coupling.
They differ, however, on whether or not they satisfy the particular OPE constraint discussed above, and 
in the shape of the vertex form factors which follow from the different models.
In spite of the different choices of these  form factors  there is, within errors, a reasonable agreement among the final results. An exception is the calculation reported in ref.~{\cite{GFW10}} using a model inspired on a Dyson--Schwinger approach which, however, as we shall see contradicts generic properties which emerge from the Constituent Chiral Quark Model which we next discuss.

\section{The Constituent Chiral Quark Model}
\nin
I have recently emphasized the need of  a simple {\it  re\-fe\-rence model} to evaluate the various hadronic contributions to $a_{\mu}$ within the same framework, and use it as a yardstick to compare with the more detailed evaluations in the literature.   
The {\it reference model} which we have proposed~{\cite{GdeR12}} is based on  
the Constituent Chiral Quark Model (C$\chi$QM) of Manohar and Georgi~{\cite{MG84}} in the presence of $SU(3)_{L}\times SU(3)_{R}$ external sources. It is an effective field theory which incorporates the interactions of  the Nambu--Goldstone modes (the low--lying pseudoscalars) of the spontaneously broken chiral symmetry, to lowest order in the chiral expansion and in the presence of chirally rotated quark fields which have become massive.
As emphasized by Weinberg~{\cite{Wei10}} the  effective Lagrangian in question is renormalizable  in the Large--${\rm N_c}$ limit and, as shown in~{\cite{EdeR11}}, the number of the required counterterms  is minimized for the choice $g_A =1$ of the axial coupling of the constituent quarks to the pseudoscalars. The model has its limitations but,
as discussed in ref.~{\cite{EdeR11}},  there is an exceptional class of  low--energy observables for which  the  predictions of the C$\chi$QM can be rather reliable. This is the case when the 
leading short--distance behaviour of the underlying Green's function of a given observable is governed by perturbative QCD. The contributions to the muon anomaly from Hadronic Vacuum Polarization, from the Hadronic Light--by--Light Scattering and from the Hadronic $Z\gamma\gamma$ vertex ( provided that $g_A =1$) fall in this class.

Figure~\ref{fig:hvpsuper} below shows the prediction of the C$\chi$QM for the HVP contribution to the muon anomaly as a function of the constituent quark mass $M_Q$ which, with $g_A =1$, is the only free parameter. 
\begin{figure}[h]
\begin{center}
\includegraphics[width=0.45\textwidth]{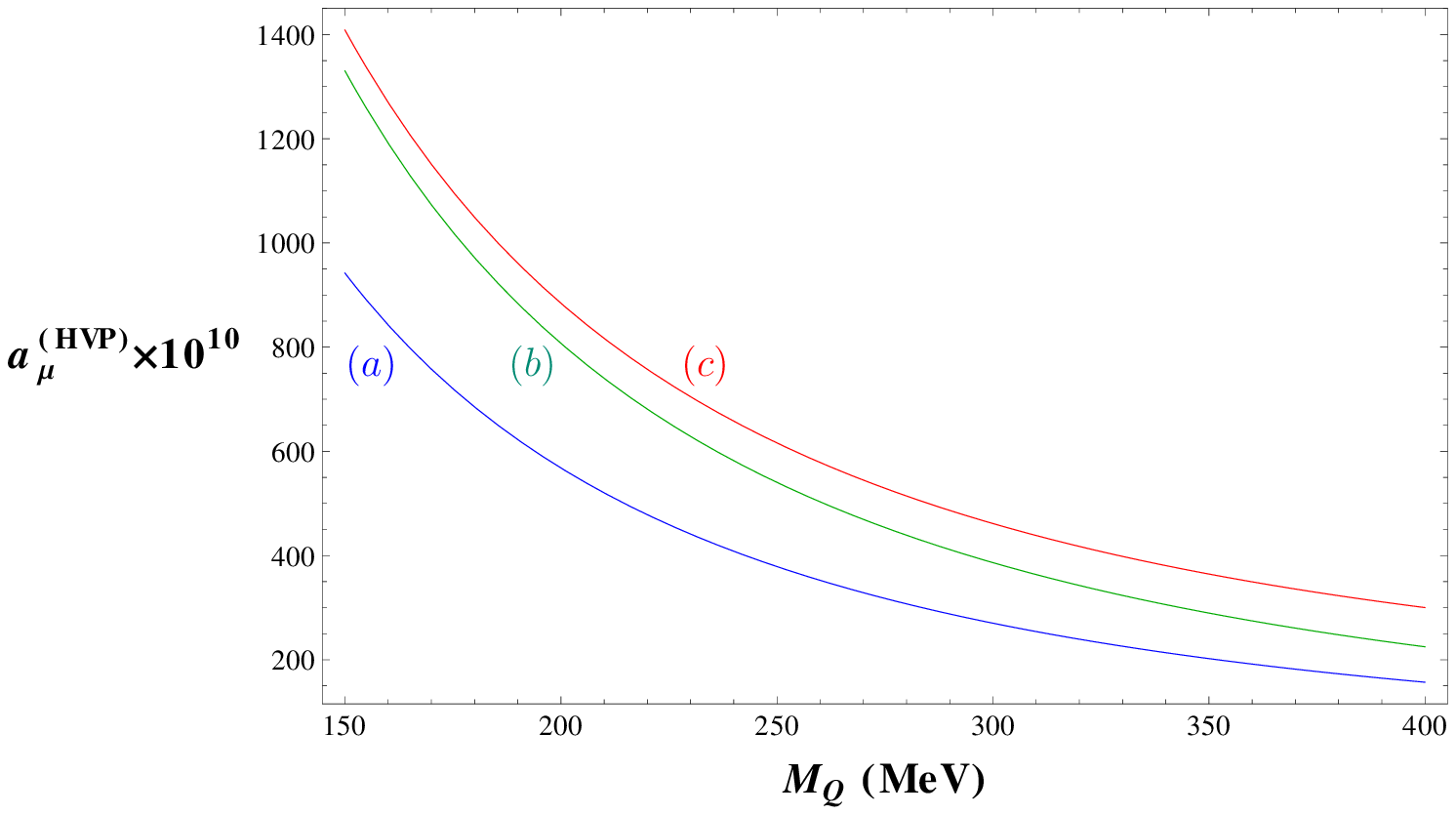}
\bf\caption{\label{fig:hvpsuper}}
{\it  The HVP Contribution in the C$\chi$QM.}
\end{center}
\end{figure}
The curve (a) in Fig.~\ref{fig:hvpsuper} is the contribution using the spectral function of the C$\chi$QM; curve (b) the same contribution as in curve (a) but with gluonic corrections included and  curve (c) the same contribution  as in curve (b) but 
with $1/\Nc$ subleading  $\pi^+ \pi^-$ and $K^+ K^-$  contributions incorporated as predicted in the C$\chi$QM. The comparison between this prediction and the phenomenological determination of $a_{\mu}^{(\rm HVP)}$ shows that fixing $M_Q$ in the range 
\beq\label{MQ}
M_Q =(240\pm 10)~\MeV\,,
\eeq
reproduces the phenomenological determination within an error of less than $10\%$.
This error, however, only reflects the phenomenological  choice that we have made  to fix $M_Q$, it does not include the systematic error of the C$\chi$QM itself. 
As shown in ref.~{\cite{GdeR12}}, with $M_Q$ fixed in this range, the higher order HVP contributions, as well as the electroweak Hadronic $Z\gamma\gamma$ contribution are reproduced rather well with the C$\chi$QM.

\begin{figure}[h]
\begin{center}
\includegraphics[width=0.45\textwidth]{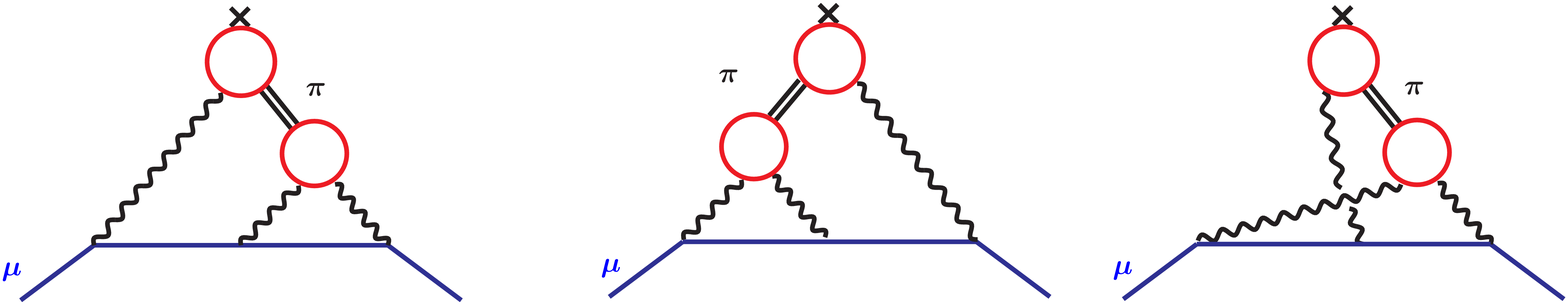}
\bf\caption{\label{fig:hlbylpiqm}}
{\it  The $\pi^0$--Exchange Contribution in the C$\chi$QM.}
\end{center}
\end{figure}

\begin{figure}[h]
\begin{center}
\includegraphics[width=0.35\textwidth]{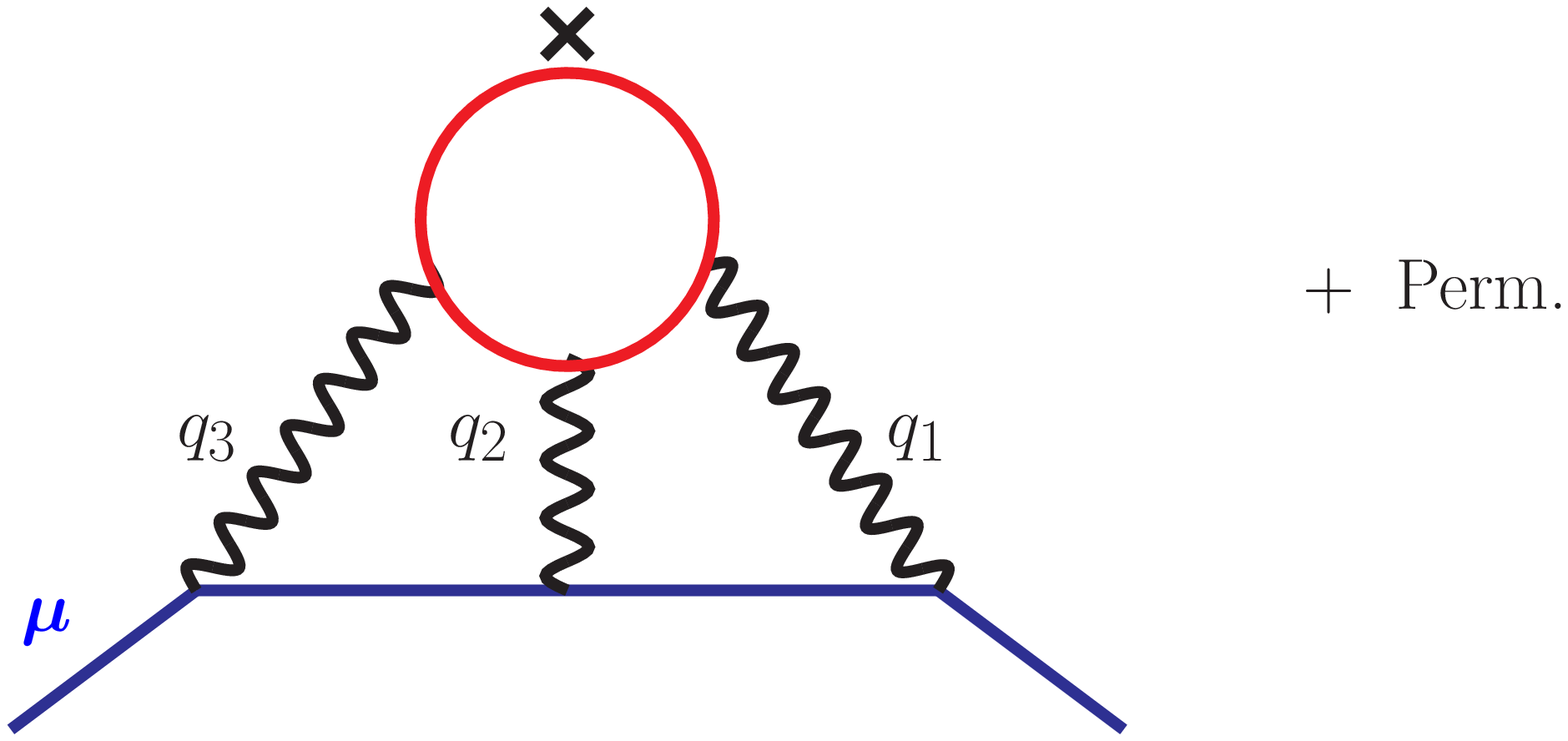}
\bf\caption{\label{fig:hlbyloop}}
{\it  The Quark Loop Contribution in the C$\chi$QM.}
\end{center}
\end{figure}

When examining the HLbyL scattering contribution to the muon anomaly in the C$\chi$QM, one finds that there are two competing contributions: one is the $\pi^0$--exchange diagrams in Fig.~\ref{fig:hlbylpiqm} where the circles there are constituent quark loops, the other one the irreducible constituent quark loop in Fig.~\ref{fig:hlbyloop}.

\nin   
An interesting feature which emerges from the calculation is the balance between the {\it Goldstone Contribution} and the {\it Quark Loop Contribution}. Indeed, as the constituent quark mass $M_Q$ gets larger and larger, the {\it Goldstone Contribution} dominates:
asymptotically, for large $M_Q$ values it  behaves (as expected) like

\begin{multline}
a^{(\rm HLbyL)}_{\mu}({\rm GC})=\left(\frac{\alpha}{\pi}\right)^{3}\Nc^2 \frac{m_{\mu}^{2}}{48\pi^2 f_{\pi}^{2}} \times \\
\Big[ \log^2\!\frac{M_{Q}}{m_{\pi}} + \cO\Big(\log\frac{M_{Q}}{m_{\pi}}\Big)\Big]\,,
\end{multline}
\nin
while for $M_Q$ smaller and smaller  it is the {\it Constituent Quark Loop Contribution} which dominates: asymptotically, for small $M_Q$ values (though still with $M_Q > m_{\mu}$) it  behaves like

{\setl
\bea
\lefteqn{a_{\mu}^{\rm HLbyL}({\rm CQL})  =  \left(\frac{\alpha}{\pi}\right)^3 \Nc \left(\sum_{q=u,d,s} Q_q^4 \right)\times}\nnb \\
 & &   
\left\{\left[\frac{3}{2}\zeta(3)-\frac{19}{16} \right] \frac{m_{\mu}^2}{M_{Q}^2}  +\cO\left(\frac{m_{\mu}^4}{M_{Q}^4}\log^2 \frac{m_{\mu}^2}{M_{Q}^2} \right)\right\}\,.
\eea}

\begin{figure}[h]
\begin{center}
\includegraphics[width=0.45\textwidth]{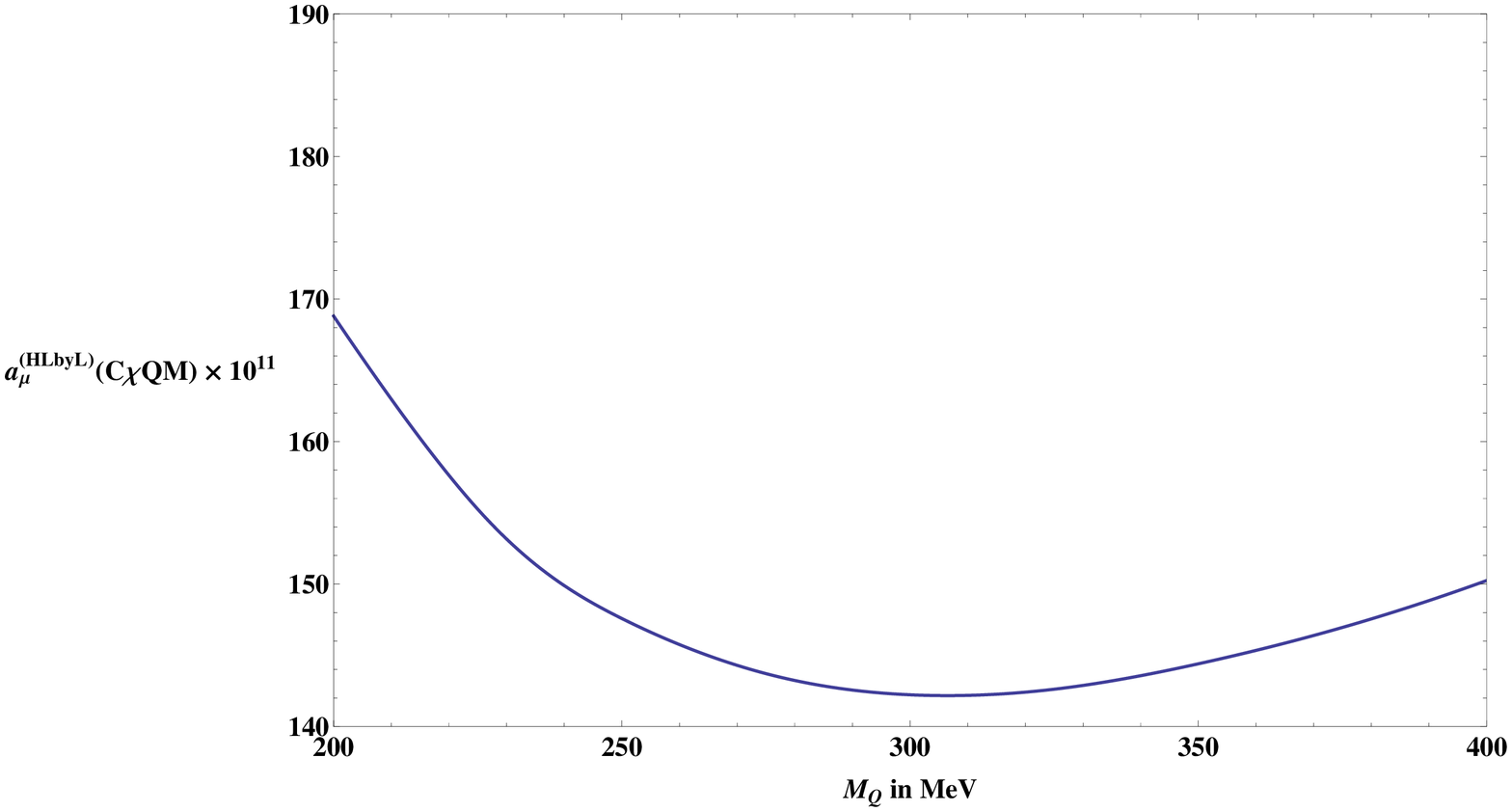}
\bf\caption{\label{fig:hlbylcqm}}
{\it  The HLbyL Contribution in the C$\chi$QM.}
\end{center}
\end{figure}

\nin
These features are illustrated by the plot of the total $a^{(\rm HLbyL)}_{\mu} ({\rm C} \chi {\rm QM})$ versus $M_Q$ shown in Fig.~\ref{fig:hlbylcqm}. In fact the plot also shows that the value of $a^{(\rm HLbyL)}_{\mu} ({\rm C} \chi {\rm QM})$ is quite stable for a rather large choice of reasonable $M_Q$ values.

The C$\chi$QM result contradicts what is reported in ref.~{\cite{GFW10}} where the equivalent contribution to the {\it constituent quark loop}  is found to be: $(136\pm 59)\times 10^{-11}$ i.e.  much larger than the contribution found by the same authors for the $\pi^0$--exchange: $(81\pm 12)\times 10^{-11}$ which, within errors, is compatible with other phenomenological determinations. This casts serious doubts about the compatibility of the model used in ref.~{\cite{GFW10}} (or perhaps of their calculation~\footnote {After the completion of this mini review, there has appeared a new version of this model in the archives~{\cite{GFW12}} with a smaller result for the quark loop contribution: $(96\pm 2)\times 10^{-11}$. Notice that the error here does not include the systematic error of the model and the calculation is claimed to be incomplete as yet.}) with basic QCD features encoded in the C$\chi$QM.

We conclude that, in the absence of more refined calculations, the number quoted in Table~2 for the light--by--light scattering contribution to the muon anomaly still represents  a valid estimate. In fact, there is a recent independent analysis in ref.~{\cite{BMZA12}} which confirms that estimate. 

\section*{Acknowledgements}
\nin
I wish to thank David Greynat for a pleasent collaboration on the C$\chi$QM calculations reported here and, together with Marc Knecht and Laurent Lellouch, for useful discussions and suggestions. 




\end{document}